\documentclass[conference, letterpaper]{IEEEtran}
\ifCLASSINFOpdf
\else
\fi
\ifCLASSINFOpdf
   \usepackage[pdftex]{graphicx}
\else
\fi

\let\labelindent\relax
\usepackage[cmex10]{amsmath}
\usepackage{color}
\usepackage{caption}
\usepackage{epstopdf}
\usepackage{tabularx}
\usepackage{pdflscape}
\usepackage{fancyhdr}
\usepackage{multirow}
\usepackage{epsfig}
\usepackage{enumitem}
\usepackage{amsmath}
\renewcommand{\thispagestyle}[2]{} 
\newcommand\mdoubleplus{\mathbin{+\mkern-10mu+}}
\fancypagestyle{plain}{
\fancyhead{}
\fancyhead[C]{first page center header}
\fancyfoot{}
\fancyfoot[C]{first page center footer}
}
\begin{document}
\title{SIT: A Lightweight Encryption Algorithm for Secure Internet of Things}
\author{\IEEEauthorblockN{Muhammad Usman\IEEEauthorrefmark{1},
		Irfan Ahmed\IEEEauthorrefmark{2},
		M. Imran Aslam\IEEEauthorrefmark{2}, Shujaat Khan\IEEEauthorrefmark{1} and Usman Ali Shah\IEEEauthorrefmark{2}}
	\IEEEauthorblockA{
		\\
		\IEEEauthorrefmark{1}Faculty of Engineering Science and Technology (FEST),\\ Iqra University, Defence View,\\ Karachi-75500, Pakistan.\\
				Email: \{musman, shujaat\}@iqra.edu.pk\\
		\\
		\IEEEauthorrefmark{2}Department of Electronic Engineering,\\ NED University of Engineering and Technology, \\University Road, Karachi 75270, Pakistan.\\
		Email: \{irfans, iaslam\}@neduet.edu.pk, uashah@gmail.com}}

\maketitle

\begin{abstract}
The Internet of Things (IoT) being a promising technology of the future is expected to connect billions of devices.  The increased number of communication is expected to generate mountains of data and the security of data can be a threat.  The devices in the architecture are essentially smaller in size and low powered.  Conventional encryption algorithms are generally computationally expensive due to their complexity and requires many rounds to encrypt, essentially wasting the constrained energy of the gadgets.  Less complex algorithm, however, may compromise the desired integrity.  In this paper we propose a lightweight encryption algorithm named as Secure IoT (SIT).  It is a 64-bit block cipher and requires 64-bit key to encrypt the data.  The architecture of the algorithm is a mixture of feistel and a uniform substitution-permutation network.  Simulations result shows the algorithm provides substantial security in just five encryption rounds.  The hardware implementation of the algorithm is done on a low cost 8-bit micro-controller and the results of code size, memory utilization and encryption/decryption execution cycles are compared with benchmark encryption algorithms. The MATLAB code for relevant simulations is available online at https://goo.gl/Uw7E0W.
\end{abstract}

\begin{IEEEkeywords}
IoT; Security; Encryption; Wireless Sensor Network WSN; Khazad
\end{IEEEkeywords}

\IEEEpeerreviewmaketitle
\section{Introduction}
\label{intro}
The Internet of Things (IoT) is turning out to be an emerging discussion in the field of research and practical implementation in the recent years.  IoT is a model that includes ordinary entities with the capability to sense and communicate with fellow devices using Internet \cite{PRIOT10}.  As the broadband Internet is now generally accessible and its cost of connectivity is also reduced, more gadgets and sensors are getting connected to it \cite{PR_IOT_22}.  Such conditions are providing suitable ground for the growth of IoT.  There is great deal of complexities around the IoT, since we wish to approach every object from anywhere in the world \cite{NIOT_12}.  The sophisticated chips and sensors are embedded in the physical things that surround us, each transmitting valuable data.  The process of sharing such large amount of data begins with the devices themselves which must securely communicate with the IoT platform.  This platform integrates the data from many devices and apply analytics to share the most valuable data with the applications.  The IoT is taking the conventional internet, sensor network and mobile network to another level as every ‘thing’ will be connected to the internet.  A matter of concern that must be kept under consideration is to ensure the issues related to confidentiality, data integrity and authenticity that will emerge on account of security and privacy \cite{PRSC2}.
\subsection{Applications of IoT:}
\label{AppIoT}
With the passage of time, more and more devices are getting connected to the Internet.  The houses are soon to be equipped with smart locks \cite{NIOT_13}, the personal computer, laptops, tablets, smart phones, smart TVs, video game consoles even the refrigerators and air conditioners have the capability to communicate over Internet.  This trend is extending outwards and  it is estimated that by the year 2020 there will be over 50 billion objects connected to the Internet \cite{NIOT_22}.  This estimates that for each person on earth there will be 6.6 objects online.  The earth will be blanketed with millions of sensors gathering information from physical objects and will upload it to the Internet. 

It is suggested  that application of IoT is yet in the early stage but is beginning to evolve rapidly \cite{PR_IOT24,PR_IOT_29}.  An overview of IoT in building automation system is given in \cite{NIOT_14}.  It is suggested in \cite{PR_IOT23} that various industries have a growing interest towards use of IoT. Various applications of IoT in health care industries are discussed in  \cite{PR_IOT_27,NIOT_19} and the improvement opportunities in health care brought in by IoT will be enormous \cite{PR_IOT_28}.

It has been predicted that IoT will contribute in the making the mining production safer \cite{PR_IOT_30} and the forecasting of disaster will be made possible.  It is expected that IoT will transform the automobile services and transportation systems \cite{PR_IOT_32}.  As more physical objects will be equipped with sensors and RFID tags transportation companies will be able to track and monitor the object movement from origin to destination \cite{PR_IOT_31}, thus IoT shows promising behavior in the logistics industry as well.

With so many applications eying to adapt the technology with the intentions to contribute in the growth of economy, health care facility, transportation and a better life style for the public, IoT must offer adequate security to their data to encourage the adaptation process. 

\subsection{Security Challenges in IoT:}
\label{SnP}
To adopt the IoT technology it is necessary to build the confidence among the users about its security and privacy that it will not cause any serious threat to their data integrity, confidentiality and authority.  Intrinsically IoT is vulnerable to various types of security threats,  if necessary security measures are not taken there will be a threat of information leakage or could prove a damage to economy  \cite{NIOT_9,shu3}.  Such threats may be considered as one of the major hindrance in IoT \cite{NIOT_20,shu4}.

IoT is extremely open to attacks \cite{NIOT_2,PRIOT19}, for the reasons that there is a fair chance of physical attack on its components as they remain unsupervised for long time.  Secondly, due to the wireless communication medium, the eavesdropping is extremely simple.  Lastly the constituents of IoT bear low competency in terms of energy with which they are operated and also in terms of computational capability.  The implementation of conventional computationally expensive security algorithms will result in the hindrance on the performance of the energy constrained devices.

It is predicted that substantial amount of data is expected to be generated while IoT is used for monitoring purposes and it is vital to preserve unification of data \cite{PRIOT21}.  Precisely, data integrity and authentication are the matters of concern. 

From a high level perspective, IoT is composed of three components namely, Hardware, Middleware and Presentation \cite{PRIOT10}.  Hardware consists of sensors and actuators, the Middleware provides storage and computing tools and the presentation provides the interpretation tools accessible on different platforms.  It is not feasible to process the data collected from billions of sensors, context-aware Middleware solutions are proposed to help a sensor decide the most important data for processing \cite{PR_IOT_26}.  Inherently the architecture of IoT does not offer sufficient margin to accomplish the necessary actions involved in the process of authentication and data integrity.  The devices in the IoT such as RFID are questionable to achieve the fundamental requirements of authentication process that includes constant communication with the servers and exchange messages with nodes.

In secure systems the confidentiality of the data is maintained and it is made sure that during the process of message exchange the data retains its originality and no alteration is unseen by the system.  The IoT is composed of many small devices such as RFIDs which remain unattended for extended times, it is easier for the adversary to access the data stored in the memory \cite{PRIOT80}.  To provide the immunity against Sybil attacks in RFID tags, received signal strength indication (RSSI) based methodologies are used in \cite{PR_IOT_33}, \cite{PR_IOT_34}, \cite{PR_IOT_35} and \cite{PR_IOT_36}.

Many solutions have been proposed for the wireless sensor networks which consider the sensor as a part of Internet connected via nodes \cite{PRIOT78}.  However, in IoT the sensor nodes themselves are considered as the Internet nodes making the authentication process even more significant.  The integrity of the data also becomes vital and requires special attention towards retaining its reliability.

\subsection{Motivation And Organization of Paper}
Recently a study by HP reveals that 70\% of the devices in IoT are vulnerable to attacks \cite{NIOT_17}.  An attack can be performed by sensing the communication between two nodes which is known as a man-in-the-middle attack. No reliable solution has been proposed to cater such attacks.  Encryption however could lead to minimize the amount of damage done to the data integrity. 
To assure data unification while it is stored on the middle ware and also during the transmission it is necessary to have a security mechanism.  Various cryptographic algorithms have been developed that addresses the said matter, but their utilization in IoT is questionable as the hardware we deal in the IoT are not suitable for the implementation of computationally expensive encryption algorithms.  A trade-off must be done to fulfil the requirement of security with low computational cost.

In this paper, we proposed a lightweight cryptographic algorithm for IoT named as Secure IoT (SIT).  The proposed algorithm is designed for IoT to deal with the security and resource utilization challenges mentioned in section \ref{SnP}.  The rest of the paper is organized as follows, in section \ref{Literature}, a short literature review is provided for the past and contemporary lightweight cryptographic algorithms, in section \ref{Proposed}, the detail architecture and functioning of the proposed algorithm is presented.  Evaluation of SIT and experimental setup is discussed in section \ref{Experiments}.  Conclusion of the paper is presented in section \ref{Conclusion}.

\section{Cryptographic Algorithms for IoT:} \label{Literature}
The need for the lightweight cryptography have been widely discussed \cite{PRLW7,shu5,shu7}, also the shortcomings of the IoT in terms of constrained devices are highlighted.  There in fact exist some lightweight cryptography algorithms that does not always exploit security-efficiency trade-offs.  Amongst the block cipher, stream cipher and hash functions, the block ciphers have shown considerably better performances.

A new block cipher named mCrypton is proposed \cite{PRLW1}. The cipher comes with the options of 64 bits, 96 bits and 128 bits key size.  The architecture of this algorithm is followed by Crypton \cite{crypton} however functions of each component is simplified to enhance its performance for the constrained hardware.  In \cite{PRLW3} the successor of Hummingbird-1 \cite{PRHUM1} is proposed as Hummingbird-2(HB-2).  With 128 bits of key and a 64 bit initialization vector Hummingbird-2 is tested to stay unaffected by all of the previously known attacks.  However the cryptanalysis of HB-2 \cite{HB2crypt} highlights the weaknesses of the algorithm and that the initial key can be recovered. \cite{PRLW10} studied different legacy encryption algorithms including RC4, IDEA and RC5 and measured their energy consumption.  They computed the computational cost of the RC4 \cite{PRLW13}, IDEA \cite{PRLW14} and RC5 ciphers on different platforms.  However, various existing algorithms were omitted during the study.

TEA \cite{TEA}, Skipjack \cite{SKJ} and RC5 algorithms have been implemented on Mica2 hardware platform \cite{PRNET1}.  To measure the energy consumption and memory utilization of the ciphers Mica2 was configured in single mote. Several block ciphers including AES \cite{AES}, XXTEA \cite{xxtea}, Skipjack and RC5 have been implemented \cite{PRLW9}, the energy consumption and execution time is measured.  The results show that in the AES algorithm the size of the key has great impact on the phases of encryption, decryption and key setup i-e the longer key size results in extended execution process. 
RC5 offers diversified parameters i-e size of the key, number of rounds and word size can be altered.  Authors have performed variety of combinations to find out that it took longer time to execute if the word size is increased.  Since key setup phase is not involved in XXTEA and Skipjack, they drew less energy but their security strength is not as much as AES and RC5.  \cite{404} proposed lightweight block cipher Simon and Speck to show optimal results in hardware and software respectively.  Both ciphers offer a range of key size and width, but atleast 22 numbers of round require to perform sufficient encryption.  Although the Simon is based on low multiplication complexity but the total number of required mathematical operation is quite high \cite{481CHK2,shu6}

\section{Proposed Algorithm}\label{Proposed}

The architecture of the proposed algorithm provides a simple structure suitable for implementing in IoT environment.  Some well known block cipher including AES (Rijndael) \cite{AES}, 3-Way \cite{3way}, Grasshopper \cite{grasshopper}, PRESENT \cite{PRESENT}, SAFER \cite{safer}, SHARK \cite{SHARK}, and Square \cite{Square} use Substitution-Permutation (SP) network.  Several alternating rounds of substitution and transposition satisfies the Shannon's confusion and diffusion properties that ensues that the cipher text is changed in a pseudo random manner.  Other popular ciphers including SF \cite{PR_SF}, Blowfish \cite{BLOWFISH}, Camelia \cite{camellia} and DES \cite{DES}, use the feistel architecture.  One of the major advantage of using feistel architecture is that the encryption and decryption operations are almost same.  The proposed algorithm is a hybrid approach based on feistel and SP networks.  Thus making use of the properties of both approaches to develop a lightweight algorithm that presents substantial security in IoT environment while keeping the computational complexity at moderate level.

SIT is a symmetric key block cipher that constitutes of 64-bit key and plain-text.  In symmetric key algorithm the encryption process consists of encryption rounds, each round is based on some mathematical functions to create confusion and diffusion.  Increase in number of rounds ensures better security but eventually results in increase in the consumption of constrained energy \cite{PR_MF_1}.  The cryptographic algorithms are usually designed to take on an average 10 to 20 rounds to keep the encryption process strong enough that suits the requirement of the system.  However the proposed algorithm is restricted to just five rounds only, to further improve the energy efficiency, each encryption round includes mathematical operations that operate on 4 bits of data.  To create sufficient confusion and diffusion of data in order to confront the attacks, the algorithm utilizes the feistel network of substitution diffusion functions.  The details of SIT design is discussed in section \ref{keyexp11} and \ref{ENCR}.

Another vital process in symmetric key algorithms is the generation of key.  The key generation process involves complex mathematical operations.  In WSN environment these operations can be performed wholly on decoder  \cite{PR_SF},\cite{SFFPGA,shu1}, on the contrary in IoT the node themselves happens to serve as the Internet node, therefore, computations involved in the process of key generation must also be reduced to the extent that it ensures necessary security.  In the sub-sections the process of key expansion and encryption are discussed in detail.  Some notations used in the explanation are shown in Table \ref{notations}
\begin{table}[!ht]
\centering
\setlength{\extrarowheight}{1pt}
\begin{tabular}{|c|c|l}
\cline{1-2}
Notation & Function &  \\ \cline{1-2}
$\bigoplus$ & XOR & \\ \cline{1-2}
$\bigodot$ & XNOR &  \\ \cline{1-2}
$\mdoubleplus$, $\parallel$ & Concatenation \\ \cline{1-2}
\end{tabular}
\captionsetup{justification=centering}
\caption{Notations}
\label{notations}
\end{table}

\subsection{\textit{Key Expansion}}\label{keyexp11}
\begin{figure}[!h]
	\begin{center}
		\centering 
		\includegraphics[width=8cm]{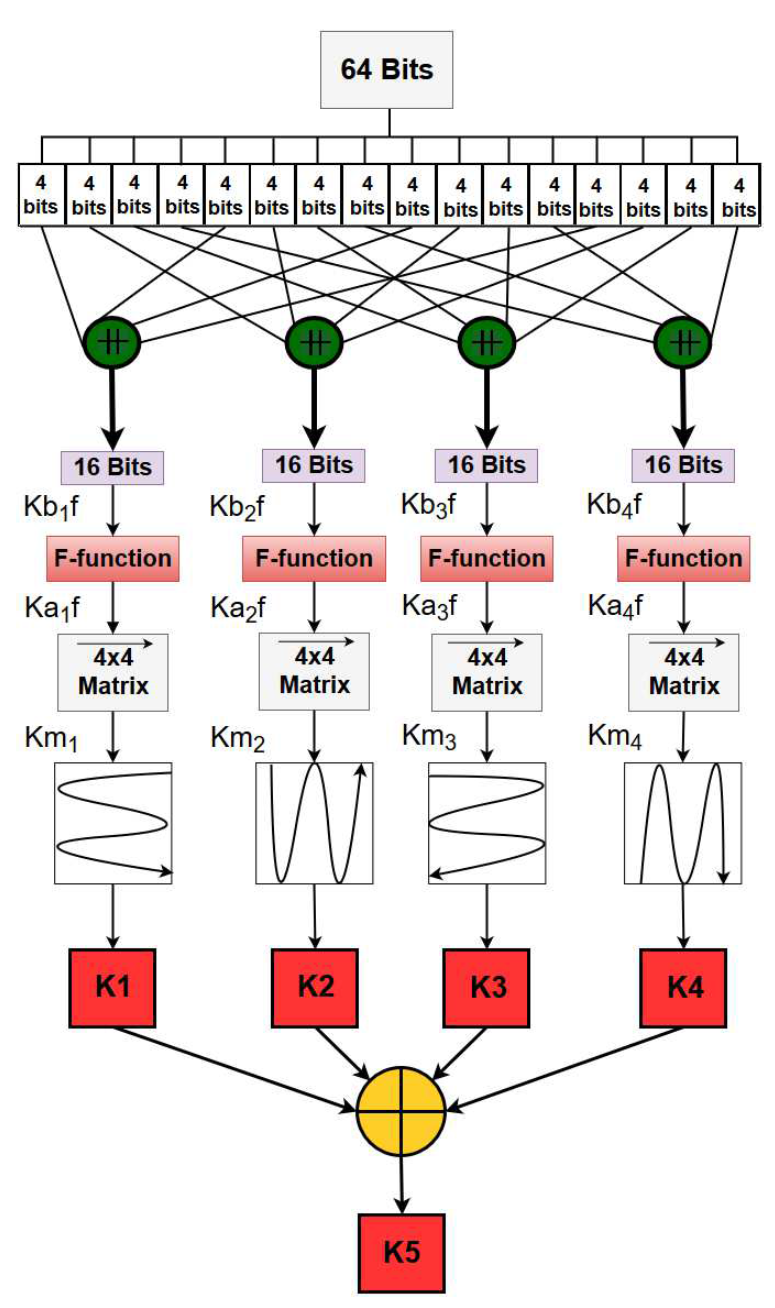} 

	\end{center}
	\caption{Key Expansion}
	\label{keyexp}
\end{figure}
The most fundamental component in the processes of encryption and decryption is the key.  It is this key on which entire security of the data is dependent, should this key be known to an attacker, the secrecy of the data is lost.  Therefore necessary measures must be taken into account to make the revelation of the key as difficult as possible.  The feistel based encryption algorithms are composed of several rounds, each round requiring a separate key.  The encryption/decryption of the proposed algorithm is composed of five rounds, therefore, we require five unique keys for the said purpose.  To do so, we introduce a key expansion block which is described in this section.

To maintain the security against exhaustive search attack the length of the true key $k_{t}$ must be large so that it becomes beyond the capability of the enemy to perform $2^{k_{t}-1}$ encryptions for key searching attacks.  The proposed algorithm is a 64-bit block cipher, which means it requires 64-bit key to encrypt 64-bits of data.  A cipher key (Kc) of 64-bits is taken as an input from the user.  This key shall serve as the input to the key expansion block.  The block upon performing substantial operations to create confusion and diffusion in the input key will generate five unique keys.  These keys shall be used in the encryption/decryption process and are strong enough to remain indistinct during attack.

The architecture of the key expansion block is shown in Fig. \ref{keyexp}.  The block uses an \textbf{\textit{f}}-function which is influenced by tweaked Khazad block cipher \cite{khazad}.  Khazad is not a feistel cipher and it follows wide trial strategy.  The wide trial strategy is composed of several linear and non-linear transformations that ensures the dependency of output bits on input bits in a complex manner \cite{WT}. Detailed explanation of the components of key expansion are discussed below:
\begin{itemize}
\item In the first step the 64-bit cipher key (Kc) is divided into the segments of 4-bits. 
\item The \textbf{\textit{f}}-function operates on 16-bits data. Therefore four \textbf{\textit{f}}-function blocks are used. These 16-bits for each \textbf{\textit{f}}-function are obtained after performing an initial substitution of segments of cipher key ($Kc$) as shown in equation (\ref{KBF}).

\begin{eqnarray}\label{KBF}
Kb_{i}f=  \parallel_{j=1}^4 Kc_{4(j-1)+i} 
\end{eqnarray}
where $i$ = 1 to 4 for first 4 round keys as shown in Fig. \ref{keyexp}.

\item The next step is to get $Ka_{i}f$ by passing the 16-bits of $Kb_{i}f$ to the \textbf{\textit{f}}-function as shown in equation (\ref{KAF}).

\begin{eqnarray}\label{KAF}
Ka_{i}f= \textbf{\textit{f}}(Kb_{i}f) 
\end{eqnarray}

\item \textbf{\textit{f}}-function is comprised of P and Q tables.  These tables perform linear and non-linear transformations resulting in confusion and diffusion as illustrated in Fig. \ref{FFunction}. 
\begin{figure}[h]
	\begin{center}
		\centering 
		\includegraphics[width=8cm]{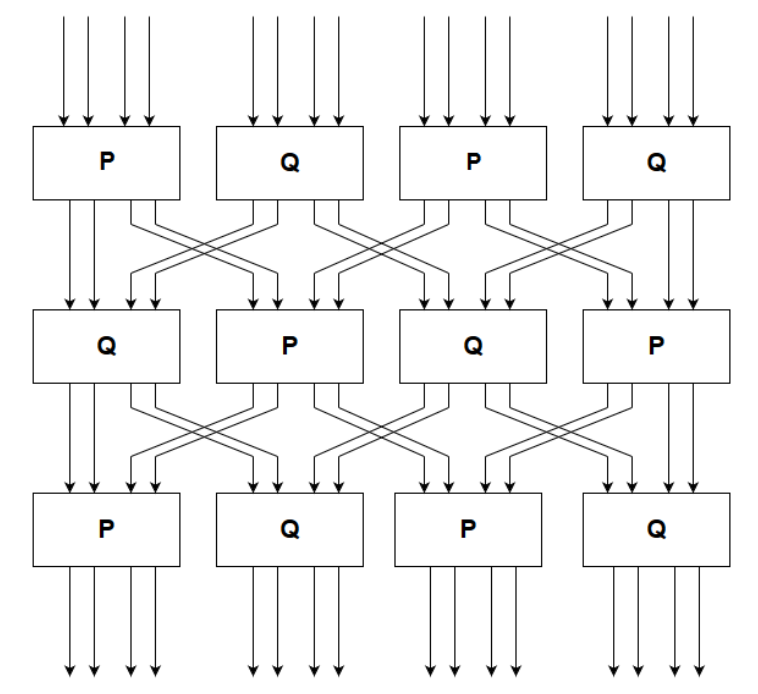}
		\end{center}
	\caption{F-Function}
	\label{FFunction}
\end{figure}
\item The transformations made by P and Q  are shown in the tables \ref{PTable} and \ref{QTable}.
\begin{table}[!h]
\centering
\begin{tabular}{|>{\centering}m{.8cm}| >{\centering}m{.02cm}| >{\centering}m{.02cm}| >{\centering}m{.02cm}| >{\centering}m{.02cm}| >{\centering}m{.02cm}| >{\centering}m{.02cm}| >{\centering}m{.02cm}| >{\centering}m{.02cm}| >{\centering}m{.02cm}| >{\centering}m{.02cm}| >{\centering}m{.02cm}| >{\centering}m{.02cm}| >{\centering}m{.02cm}| >{\centering}m{.02cm}| >{\centering\arraybackslash}m{.02cm}|>{\centering\arraybackslash}m{.02cm}|}
\hline
\textit{$ Kc{i} $} & \textbf{0} & \textbf{1} & \textbf{2} & \textbf{3} & \textbf{4} & \textbf{5} & \textbf{6} & \textbf{7} & \textbf{8} & \textbf{9} & \textbf{A} & \textbf{B} & \textbf{C} & \textbf{D} & \textbf{E} & \textbf{F}   \\ \hline
\textit{P($ Kc_{i} $)} &3 & F & E & 0 & 5 & 4 & B & C & D & A & 9 & 6 & 7 & 8 & 2 & 1 \\ \hline
\end{tabular}
\captionsetup{justification=centering}
\caption{P Table}
\label{PTable}
\end{table}

\begin{table}[]
\centering
\begin{tabular}{|>{\centering}m{.8cm}| >{\centering}m{.02cm}| >{\centering}m{.02cm}| >{\centering}m{.02cm}| >{\centering}m{.02cm}| >{\centering}m{.02cm}| >{\centering}m{.02cm}| >{\centering}m{.02cm}| >{\centering}m{.02cm}| >{\centering}m{.02cm}| >{\centering}m{.02cm}| >{\centering}m{.02cm}| >{\centering}m{.02cm}| >{\centering}m{.02cm}| >{\centering}m{.02cm}| >{\centering\arraybackslash}m{.02cm}|>{\centering\arraybackslash}m{.02cm}|}
\hline
\textit{$ Kc{i} $} & \textbf{0} & \textbf{1} & \textbf{2} & \textbf{3} & \textbf{4} & \textbf{5} & \textbf{6} & \textbf{7} & \textbf{8} & \textbf{9} & \textbf{A} & \textbf{B} & \textbf{C} & \textbf{D} & \textbf{E} & \textbf{F}   \\ \hline
\textit{Q($ Kc_{i} $)} &9 & E & 5 & 6 & A & 2 & 3 & C & F & 0 & 4 & D & 7 & B & 1 & 8 \\ \hline
\end{tabular}
\captionsetup{justification=centering}
\caption{Q Table}
\label{QTable}
\end{table}
\item The output of each \textbf{\textit{f}}-function is arranged in $4 \times 4$ matrix named $Km$ shown below:
\begin{eqnarray}
Km_{1}=
 \begin{bmatrix}

    Ka_{1}f_{1} & Ka_{1}f_{2}  & Ka_{1}f_{3}  & Ka_{1}f_{4}  \\
    Ka_{1}f_{5} & Ka_{1}f_{6}  & Ka_{1}f_{7}  & Ka_{1}f_{8} \\
    Ka_{1}f_{9} & Ka_{1}f_{10}  & Ka_{1}f_{11}  & Ka_{1}f_{12} \\
    Ka_{1}f_{13} & Ka_{1}f_{14}  & Ka_{1}f_{15}  & Ka_{1}f_{16}
  \end{bmatrix}
\end{eqnarray}
\begin{eqnarray}
Km_{2}=
  \begin{bmatrix}
    Ka_{2}f_{1} & Ka_{2}f_{2}  & Ka_{2}f_{3}  & Ka_{2}f_{4}  \\
    Ka_{2}f_{5} & Ka_{2}f_{6}  & Ka_{2}f_{7}  & Ka_{2}f_{8} \\
    Ka_{2}f_{9} & Ka_{2}f_{10}  & Ka_{2}f_{11}  & Ka_{2}f_{12} \\
    Ka_{2}f_{13} & Ka_{2}f_{14}  & Ka_{2}f_{15}  & Ka_{2}f_{16}
  \end{bmatrix}
\end{eqnarray}

\begin{eqnarray}
Km_{3}=
  \begin{bmatrix}
    Ka_{3}f_{1} & Ka_{3}f_{2}  & Ka_{3}f_{3}  & Ka_{3}f_{4}  \\
    Ka_{3}f_{5} & Ka_{3}f_{6}  & Ka_{3}f_{7}  & Ka_{3}f_{8} \\
    Ka_{3}f_{9} & Ka_{3}f_{10}  & Ka_{3}f_{11}  & Ka_{3}f_{12} \\
    Ka_{3}f_{13} & Ka_{3}f_{14}  & Ka_{3}f_{15}  & Ka_{3}f_{16}
  \end{bmatrix}
\end{eqnarray}

\begin{eqnarray}
Km_{4}=
  \begin{bmatrix}
    Ka_{4}f_{1} & Ka_{4}f_{2}  & Ka_{4}f_{3}  & Ka_{4}f_{4}  \\
    Ka_{4}f_{5} & Ka_{4}f_{6}  & Ka_{4}f_{7}  & Ka_{4}f_{8} \\
    Ka_{4}f_{9} & Ka_{4}f_{10}  & Ka_{4}f_{11}  & Ka_{4}f_{12} \\
    Ka_{4}f_{13} & Ka_{4}f_{14}  & Ka_{4}f_{15}  & Ka_{4}f_{16}
  \end{bmatrix}
\end{eqnarray}
\item To obtain round keys, K1, K2, K3 and K4 the matrices are transformed into four arrays of 16 bits that we call round keys (Kr). The arrangement of these bits are shown in equations (\ref{k1}), (\ref{k2}), (\ref{k3}) and (\ref{k4}).
\begin{multline}\label{k1}
K1 = a_{4} \mdoubleplus a_{3} \mdoubleplus a_{2} \mdoubleplus a_{1}\mdoubleplus a_{5} \mdoubleplus a_{6} \mdoubleplus a_{7} \mdoubleplus a_{8}\\\mdoubleplus a_{12} \mdoubleplus a_{11} \mdoubleplus a_{10}\mdoubleplus a_{9} \mdoubleplus a_{13} \mdoubleplus a_{14} \mdoubleplus a_{15} \mdoubleplus a_{16}   
\end{multline}

\begin{multline}\label{k2}
K2 =  b_{1} \mdoubleplus b_{5} \mdoubleplus b_{9} \mdoubleplus b_{13}\mdoubleplus b_{14} \mdoubleplus b_{10} \mdoubleplus b_{6} \mdoubleplus b_{2}\\\mdoubleplus b_{3} \mdoubleplus b_{7} \mdoubleplus b_{11}\mdoubleplus b_{15} \mdoubleplus b_{16} \mdoubleplus b_{12} \mdoubleplus b_{8} \mdoubleplus b_{4} 
\end{multline}

\begin{multline}\label{k3}
K3 =  c_{1} \mdoubleplus c_{2} \mdoubleplus c_{3} \mdoubleplus c_{4}\mdoubleplus c_{8} \mdoubleplus c_{7} \mdoubleplus c_{6} \mdoubleplus c_{5}\\ \mdoubleplus c_{9} \mdoubleplus c_{10} \mdoubleplus c_{11}\mdoubleplus c_{12} \mdoubleplus c_{16} \mdoubleplus c_{15} \mdoubleplus c_{14} \mdoubleplus c_{13}
\end{multline}   

\begin{multline}\label{k4}
K4 =  d_{13} \mdoubleplus d_{9} \mdoubleplus d_{5} \mdoubleplus d_{1}\mdoubleplus d_{2} \mdoubleplus d_{6} \mdoubleplus d_{10} \mdoubleplus d_{14}\\ \mdoubleplus d_{15} \mdoubleplus d_{11} \mdoubleplus d_{7}\mdoubleplus d_{3} \mdoubleplus d_{4} \mdoubleplus d_{8} \mdoubleplus d_{12} \mdoubleplus d_{16}   
\end{multline}

\item An \textit{XOR} operation is performed among the four round keys to obtain the fifth key as shown in equation (\ref{K5}).
\begin{eqnarray}\label{K5}
K5 = \bigoplus_{i=1}^{4} K{i} 
\end{eqnarray}
\end{itemize}

\subsection{\textit{Encryption}}\label{ENCR}
\begin{figure}[!ht]
	\begin{center}
		\centering 
		\includegraphics[width=8cm]{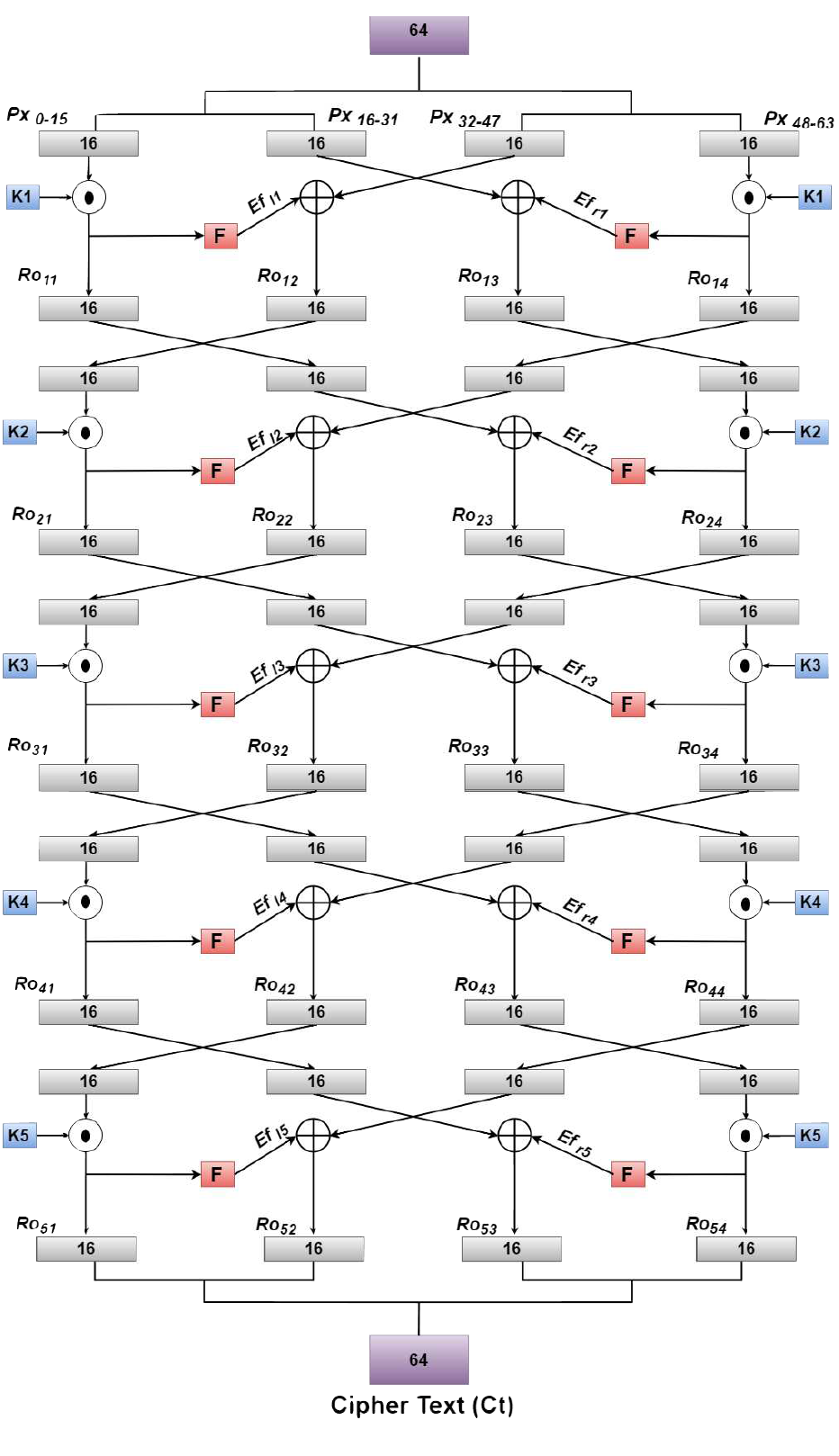}
	\end{center}
	\caption{Encryption Process}
	\label{fig:encrypt}
\end{figure}

After the generation of round keys the encryption process can be started.  For the purpose of creating confusion and diffusion this process is composed of some logical operations, left shifting, swapping and substitution.  The process of encryption is illustrated in Fig. \ref{fig:encrypt}.
For the first round an array of 64 bit plain text (Pt) is first furcated into four segments of 16 bits \textbf{$Px_{0-15}$}, \textbf{$Px_{16-31}$}, \textbf{$Px_{32-47}$} and \textbf{$Px_{48-63}$}.  As the bits progresses in each round the swapping operation is applied so as to diminish the data originality by altering the order of bits, essentially increasing confusion in cipher text.  Bitwise \textit{XNOR} operation is performed between the respective round key $K_{i}$ obtained earlier from key expansion process and \textbf{$Px_{0-15}$} and the same is applied between $K_{i}$ and \textbf{$Px_{48-63}$} resulting in $Ro_{11}$  and $Ro_{14}$ respectively.  The output of \textit{XNOR} operation is then fed to the \textbf{\textit{f}}-function generating the result $Ef_{l1}$ and $Ef_{r1}$ as shown in Fig. \ref{keyexp}. 

The \textbf{\textit{f}}-function used in encryption is the same as of key expansion, comprised of swapping and substitution operations the details of which are discussed earlier in section \ref{keyexp11}.  Bitwise \textit{XOR} function is applied between \textbf{$Ef_{l1}$} \& \textbf{$Px_{32-47}$} to obtain $Ro_{12}$ and \textbf{$Ef_{r1}$} \& \textbf{$Px_{16-31}$} to obtain $Ro_{13}$. 
\\
\begin{eqnarray}\label{Round_Enc}
Ro_{i,j} =\left\{\begin{matrix}
Px_{i,j}\bigodot  K_i \quad ; \quad j=1 \& 4& \\ 
Px_{i,j+1} \bigoplus Ef_{li} \quad ;\quad  j=2& \\ 
Px_{i,j-1} \bigoplus Ef_{ri} \quad ;\quad  j=3 & 
\end{matrix}\right.
\end{eqnarray}
Finally a round transformation is made in such a way that for succeeding round $Ro_{11}$ will become \textbf{$Px_{16-31}$}, $Ro_{12}$ will become \textbf{$Px_{0-15}$}, $Ro_{13}$ will become \textbf{$Px_{48-63}$} and $Ro_{13}$ will become \textbf{$Px_{32-47}$} as shown in Fig. \ref{fig:encrypt}.
\\

Same steps are repeated for the remaining rounds using equation (\ref{Round_Enc}).  The results of final round are concatenated to obtain Cipher Text (Ct) as shown in equation (\ref{CTT}).

\begin{eqnarray}\label{CTT}
Ct= R_{51}\mdoubleplus R_{52}\mdoubleplus R_{53}\mdoubleplus R_{54} 
\end{eqnarray}

\section{Security Analysis}
The purpose of a cipher is to provide protection to the plaintext.  The attacker intercepts the ciphertext and tries to recover the plain text.  A cipher is considered to be broken if the enemy is able to determine the secret key.  If the attacker can frequently decrypt the ciphertext without determining the secret key, the cipher is said to be partially broken.  We assume that the enemy has complete access of what is being transmitted through the channel.  The attacker may have some additional information as well but to assess the security of a cipher, the computation capability of the attacker must also be considered.  

Since the proposed algorithm is a combination of feistel and uniform substitution -combination network, it benefits from existing security analysis.  In the following a the existing security analysis of these two primitives are recalled and their relevancy with the proposed algorithm is discussed.  

\subsection{Linear and Differential Cryptanalysis}
The \textbf{\textit{f}}-function is inspired by \cite{khazad} whose cryptanalysis shows that differential and linear attacks does not have the succeed for complete cipher.  The input and output correlation is very large if the linear approximation is done for two rounds.  Also the round transformation is kept uniform which treats every bit in a similar manner and provides opposition to differential attacks. 

\subsection{Weak Keys}
The ciphers in which the non-linear operations depend on the actual key value maps the block cipher with detectable weakness.  Such case occurs in \cite{WT}.  However proposed algorithm does not use the actual key in the cipher, instead the is first \textit{XORed} and then fed to the \textbf{\textit{f}}-function.  In the \textbf{\textit{f}}-function all the non-linearity is fixed and there is no limitation on the selection of key.

\subsection{Related Keys}
An attack can be made by performing cipher operations using unknown or partially known keys.  The related key attack mostly relies upon either slow diffusion or having symmetry in key expansion block.  The key expansion process of proposed algorithm is designed for fast and non-linear diffusion of cipher key difference to that of round keys.

\subsection{Interpolation Attacks}
These attacks are dependent upon the simple structures of the cipher components that may yield a  rational expression with  a handy complexity.  The expression of the S-box of the proposed algorithm along with the diffusion layer makes such type of attack impracticable.

\subsection{SQUARE Attack}
This attack was presented by \cite{khazad} to realize how efficiently the algorithm performs against it.  The attack is able to recover one byte of the last key and the rest of keys can be recovered by repeating the attack eight times.  However to be able to do so, the attack requires $2^{8}$ key guesses by  $2^{8}$ plaintexts which is equal to $2^{16}$ S-box lookups.

\section{Experimental Setup}\label{Experiments}
\subsection{Evaluation Parameters}

To test the security strength of the proposed algorithm, the algorithm is evaluated on the basis of the following criterion.  Key sensitivity, effect of cipher on the entropy, histogram and correlation of the image.  We further tested the algorithm for computational resource utilization and computational complexity.  For this we observe the memory utilization and total computational time utilized by the algorithm for the key generation, encryption and decryption.

\subsubsection{Key Sensitivity}

An encryption algorithm must be sensitive to the key. It means that the algorithm must not retrieve the original data if the key has even a minute difference from the original key.  Avalanche test is used to evaluate the amount of alterations occurred in the cipher text by changing one bit of the key or plain text.  According to Strict Avalanche Criterion SAC \cite{PE6} if 50\% of the bits are changed due to one bit change, the test is considered to be perfect.  To visually observe this effect, we decrypt the image with a key that has a difference of only one bit from the correct key.

\subsubsection{Execution Time}

One of the fundamental parameter for the evaluation of the algorithm is the amount of time it takes to encode and decode a particular data. The proposed algorithm is designed for the IoT environment must consume minimal time and offer considerable security.

\subsubsection{Memory Utilization}

Memory utilization is a major concern in resource constrain IoT devices. An encryption algorithm is composed of several computational rounds that may occupy significant memory making it unsuitable to be utilized in IoT.  Therefore the proposed algorithm is evaluated in terms of its memory utilization.  Smaller amount of memory engagement will be favourable for its deployment in IoT.

\subsubsection{Image Histogram}

A method to observe visual effect of the cipher is to encrypt an image with the proposed algorithm and observe the randomness it produces in the image.  To evaluate the generated randomness, histogram of the image is calculated.  A uniform histogram after encryption depicts appreciable security.

\subsubsection{Image Entropy}

The encryption algorithm adds extra information to the data so as to make it difficult for the intruder to differentiate between the original information and the one added by the algorithm.  We measure the amount of information in terms of entropy, therefore it can be said that higher the entropy better is the performance of security algorithm.  To measure the entropy (H) for an image, equation (\ref{entropy}) is applied on the intensity (I) values $P(I_{i})$ being the probability of intensity value $I_{i}$.  

\begin{eqnarray}
\label{entropy}
H(I)=-\sum\limits_{i=1}^{2^{8}}P(I_{i}) \log_{b} P(I_{i})
\end{eqnarray}

\subsubsection{Correlation}\label{Corel}
The correlation between two values is a statistical relationship that depicts the dependency of one value on another.  Data points that hold substantial dependency has a significant correlation value.  A good cipher is expected to remove the dependency of the cipher text from the original message.  Therefore no information can be extracted from the cipher alone and no relationship can be drawn between the plain text and cipher text.  This criterion is best explained by Shannon in his communication theory of secrecy systems \cite{shan}.  

In this experiment we calculated the correlation coefficient for original and encrypted images.  The correlation coefficient $\gamma$ is calculated using equation (\ref{Corrcoff}).  For ideal cipher case $\gamma$ should be equal to $0$ and for the worst case $\gamma$ will be equal to $1$.

\begin{eqnarray}
\label{Corrcoff}
\gamma_{x,y}=\frac{cov(x,y)}{\sqrt{D(x)\sqrt{D(y)}}}, &	with &D(x)
\end{eqnarray}

where $cov(x,y)$, $D(x)$ and $D(y)$ are covariance and variances of variable $x$ and $y$ respectively.  The spread of values or variance of any single dimension random variable can be calculated using equation (\ref{Variance}).  Where $D(x)$ is the variance of variable $x$.

\begin{eqnarray}
\label{Variance}
D(x)=\frac{1}{N}\sum\limits_{i=1}^{N}(x_{i}-E(x))^2,
\end{eqnarray}

For the covariance between two random variables the equation (\ref{Variance}) can be transformed into equation (\ref{Covariance}).  Where $cov(x,y)$ is the covariance between two random variables $x$ and $y$.

\begin{eqnarray}
\label{Covariance}
cov(x,y)=\frac{1}{N}\sum\limits_{i=1}^{N}(x_{i}-E(x))(y_{i}-E(y)),
\end{eqnarray}

In equation (\ref{Variance}) and (\ref{Covariance}) $E(x)$ and $E(y)$ are the expected values of variable $x$ and $y$.  The expectation can be calculated using equation (\ref{Expectation}).
 
\begin{eqnarray}
\label{Expectation}
E(x)=\frac{1}{N}\sum\limits_{i=1}^{N}x_{i},
\end{eqnarray}
where $N$ is the total pixels of the image, $N=row\times col$, $x$ is a vector of length $N$ and $x_{i}$ is the ${i}$th intensity values of the original image.

\subsection{Results}
The simulation of the algorithm is done to perform the standard tests including Avalanche and image entropy and histogram on Intel Core i7-3770@3.40 GHz processor using MATLAB\textregistered.  To evaluate the performance in the real IoT environment we implemented the algorithm on ATmega 328 based Ardinuo Uni board as well.  The memory utilization and execution time of the proposed algorithm is observed.  The execution time is found to be 0.188 milliseconds and 0.187 milliseconds for encryption and decryption respectively, the proposed algorithm utilizes the 22 bytes of memory on ATmega 328 platform.  We compare our algorithm with other algorithms being implemented on hardware as shown in table \ref{comp}.

\begin{table}[!h]

\begin{tabular}{>{\centering}m{.9cm}| >{\centering}m{1cm}| >{\centering}m{.5cm}| >{\centering}m{.5cm}| >{\centering}m{.5cm}| >{\centering}m{.5cm}| >{\centering}m{.5cm}| >{\centering}m{.5cm} }

	\textbf{CIPHER} & \textbf{DEVICE} & \textbf{Block Size} & \textbf{Key Size}  & \textbf{Code Size}  & \textbf{RAM}  & \textbf{Cycles (enc)} & \textbf{Cycles (dec)}
	\end{tabular}
	\begin{tabular}{>{\centering}m{.9cm}| >{\centering}m{1cm}| >{\centering}m{.5cm}| >{\centering}m{.5cm}| >{\centering}m{.5cm}| >{\centering}m{.5cm}| >{\centering}m{.5cm}| >{\centering}m{.5cm} }
	\hline
	 AES \cite{T1} & AVR & 64 & 128 & 1570 & - & 2739 & 3579 \\ 
	 \end{tabular}
	 \begin{tabular}{>{\centering}m{.9cm}| >{\centering}m{1cm}| >{\centering}m{.5cm}| >{\centering}m{.5cm}| >{\centering}m{.5cm}| >{\centering}m{.5cm}| >{\centering}m{.5cm}| >{\centering}m{.5cm} }
	 	\hline
	 HIGHT (\cite{T2})  & AVR & 64 & 128 & 5672 & - & 2964 & 2964 \\ 
	 \end{tabular}
		 \begin{tabular}{>{\centering}m{.9cm}| >{\centering}m{1cm}| >{\centering}m{.5cm}| >{\centering}m{.5cm}| >{\centering}m{.5cm}| >{\centering}m{.5cm}| >{\centering}m{.5cm}| >{\centering}m{.5cm} }
		 	\hline
	 IDEA (\cite{T5})  & AVR & 64 & 80 & 596 & - & 2700 & 15393 \\ 
	  \end{tabular}
	 	 \begin{tabular}{>{\centering}m{.9cm}| >{\centering}m{1cm}| >{\centering}m{.5cm}| >{\centering}m{.5cm}| >{\centering}m{.5cm}| >{\centering}m{.5cm}| >{\centering}m{.5cm}| >{\centering}m{.5cm} }
	 	 	\hline
	 KATAN (\cite{T5})& AVR & 64 & 80 & 338 & 18 & 72063 & 88525 \\ 
	  \end{tabular}
	 	 \begin{tabular}{>{\centering}m{.9cm}| >{\centering}m{1cm}| >{\centering}m{.5cm}| >{\centering}m{.5cm}| >{\centering}m{.5cm}| >{\centering}m{.5cm}| >{\centering}m{.5cm}| >{\centering}m{.5cm} }
	 	 	\hline
	 KLEIN (\cite{T5})& AVR & 64 & 80 & 1268 & 18 & 6095 & 7658 \\ 
	  \end{tabular}
	 	 \begin{tabular}{>{\centering}m{.9cm}| >{\centering}m{1cm}| >{\centering}m{.5cm}| >{\centering}m{.5cm}| >{\centering}m{.5cm}| >{\centering}m{.5cm}| >{\centering}m{.5cm}| >{\centering}m{.5cm} }
	 	 	\hline
	 PRESENT (\cite{T5})& AVR & 64 & 128 & 1000 & 18 & 11342 & 13599 \\ 
	  \end{tabular}
	 	 \begin{tabular}{>{\centering}m{.9cm}| >{\centering}m{1cm}| >{\centering}m{.5cm}| >{\centering}m{.5cm}| >{\centering}m{.5cm}| >{\centering}m{.5cm}| >{\centering}m{.5cm}| >{\centering}m{.5cm} }
	 	 	\hline
	TEA (\cite{T5}) & AVR & 64 & 128 & 648 & 24 & 7408 & 7539 \\
	 \end{tabular}
		 \begin{tabular}{>{\centering}m{.9cm}| >{\centering}m{1cm}| >{\centering}m{.5cm}| >{\centering}m{.5cm}| >{\centering}m{.5cm}| >{\centering}m{.5cm}| >{\centering}m{.5cm}| >{\centering}m{.5cm} }
		 	\hline
	 PRINCE (\cite{T6}) & AVR & 64 & 128 & 1574 & 24 & 3253 & 3293 \\ 
	  \end{tabular}
	 	 \begin{tabular}{>{\centering}m{.9cm}| >{\centering}m{1cm}| >{\centering}m{.5cm}| >{\centering}m{.5cm}| >{\centering}m{.5cm}| >{\centering}m{.5cm}| >{\centering}m{.5cm}| >{\centering}m{.5cm} }
	 	 	\hline
	 SKIPJACK (\cite{T3}& Power TOSSIM & 64 & 80 & 5230 & 328 & 17390 & - \\ 
	  \end{tabular}
	 	 \begin{tabular}{>{\centering}m{.9cm}| >{\centering}m{1cm}| >{\centering}m{.5cm}| >{\centering}m{.5cm}| >{\centering}m{.5cm}| >{\centering}m{.5cm}| >{\centering}m{.5cm}| >{\centering}m{.5cm} }
	 	 	\hline
	 RC5 (\cite{T3} & Power TOSSIM & 64 & 128 & 3288 & 72 & 70700 & - \\
	  \end{tabular}
	 	 \begin{tabular}{>{\centering}m{.9cm}| >{\centering}m{1cm}| >{\centering}m{.5cm}| >{\centering}m{.5cm}| >{\centering}m{.5cm}| >{\centering}m{.5cm}| >{\centering}m{.5cm}| >{\centering}m{.5cm} }
	 	 	\hline 
	 \textbf{SIT} & ATmega328 & 64 & 64 & 826 & 22 & 3006 & 2984 \\
	  \end{tabular}

\caption{Results for Hardware Implementations}
\label{comp}	 

\end{table}

Block and key size is in bits while code and RAM is in bytes.  The cycles include key expansions along with encryption and decryption.

\begin{figure}[!h]

	\begin{center}
		\begin{tabular}{>{\centering}m{1.42cm} >{\centering}m{1.42cm} >{\centering\arraybackslash}m{1.42cm}>{\centering\arraybackslash}m{1.42cm}}
			  Orignal Image & Encrypted Image & Decrypted with correct key & Decrypted with  wrong key \\
			
				\raisebox{-\totalheight}{\centering \fbox{\includegraphics[width=1.5cm]{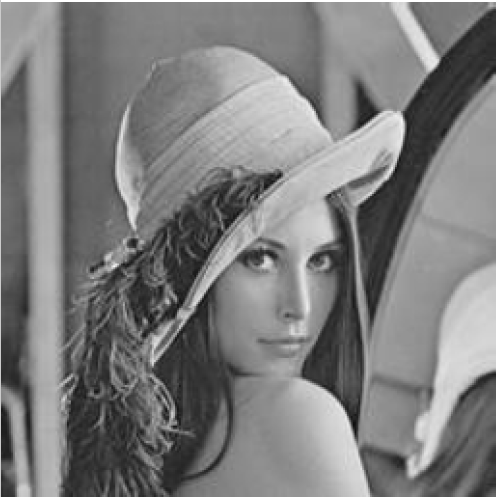}}}
			& 
			\raisebox{-\totalheight}{\centering \fbox{\includegraphics[width=1.5cm]{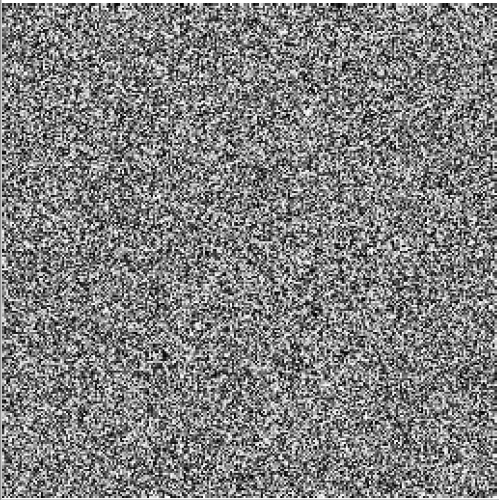}}}
			& 
			\raisebox{-\totalheight}{\centering \fbox{\includegraphics[width=1.5cm]{Orignal_Lena.PNG}}}
			& 
			\raisebox{-\totalheight}{\centering \fbox{\includegraphics[width=1.5cm]{Encrypted_Lena.PNG}}}
			\\
			\raisebox{-\totalheight}{\centering \fbox{\includegraphics[width=1.5cm]{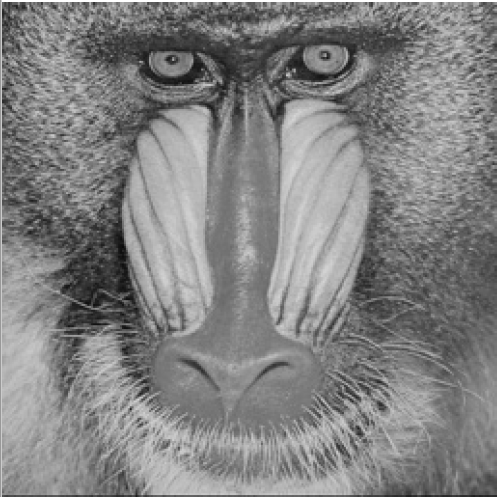}}}
			& 
			\raisebox{-\totalheight}{\centering \fbox{\includegraphics[width=1.5cm]{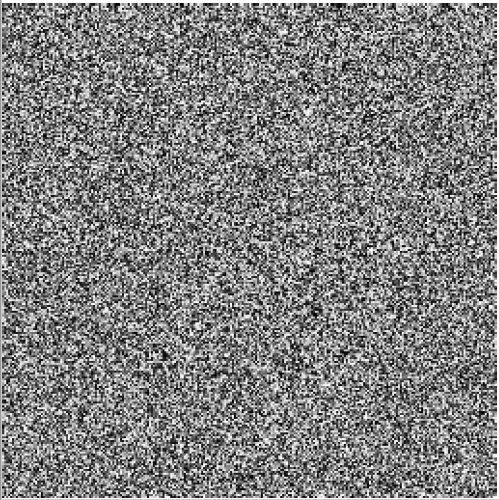}}}
			& 
			\raisebox{-\totalheight}{\centering \fbox{\includegraphics[width=1.5cm]{Orignal_baboon.PNG}}}
			& 
			\raisebox{-\totalheight}{\centering \fbox{\includegraphics[width=1.5cm]{Encrypted_baboon.PNG}}}
			\\
			\raisebox{-\totalheight}{\centering \fbox{\includegraphics[width=1.5cm]{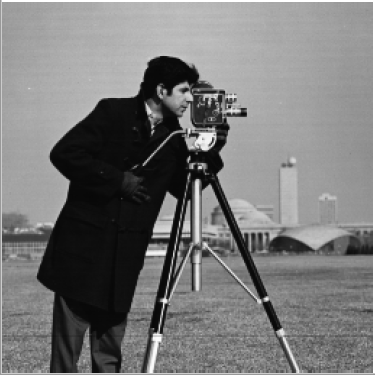}}}
			& 
			\raisebox{-\totalheight}{\centering \fbox{\includegraphics[width=1.5cm]{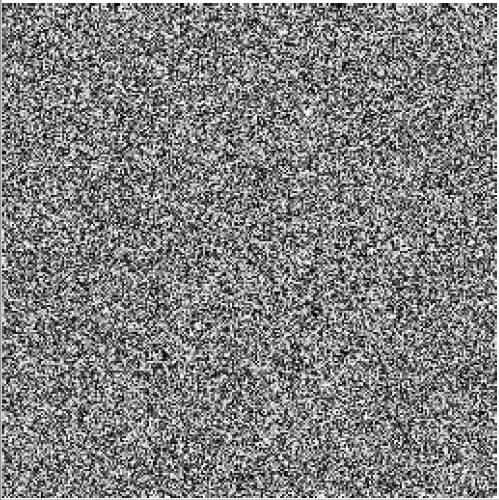}}}
			& 
			\raisebox{-\totalheight}{\centering \fbox{\includegraphics[width=1.5cm]{Orignal_cameraman.PNG}}}
			& 
			\raisebox{-\totalheight}{\centering \fbox{\includegraphics[width=1.5cm]{Encrypted_cameraman.PNG}}}			
			\\
			 \raisebox{-\totalheight}{\centering \fbox{\includegraphics[width=1.5cm]{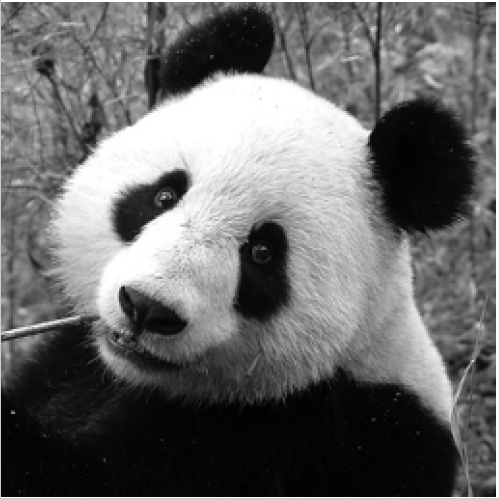}}}
			& 
			\raisebox{-\totalheight}{\centering \fbox{\includegraphics[width=1.5cm]{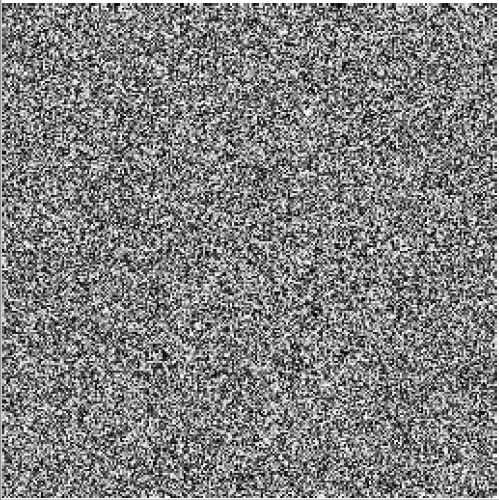}}}
			& 
			\raisebox{-\totalheight}{\centering \fbox{\includegraphics[width=1.5cm]{Orignal_panda.PNG}}}
			& 
			\raisebox{-\totalheight}{\centering \fbox{\includegraphics[width=1.5cm]{Encrypted_panda.PNG}}}
			\\
		\end{tabular}
		\caption{Image decryption and key sensitivity}
		\label{Images1}
\end{center}
\end{figure}

\begin{figure}[!ht]

	\begin{center}
			\includegraphics[width=8cm]{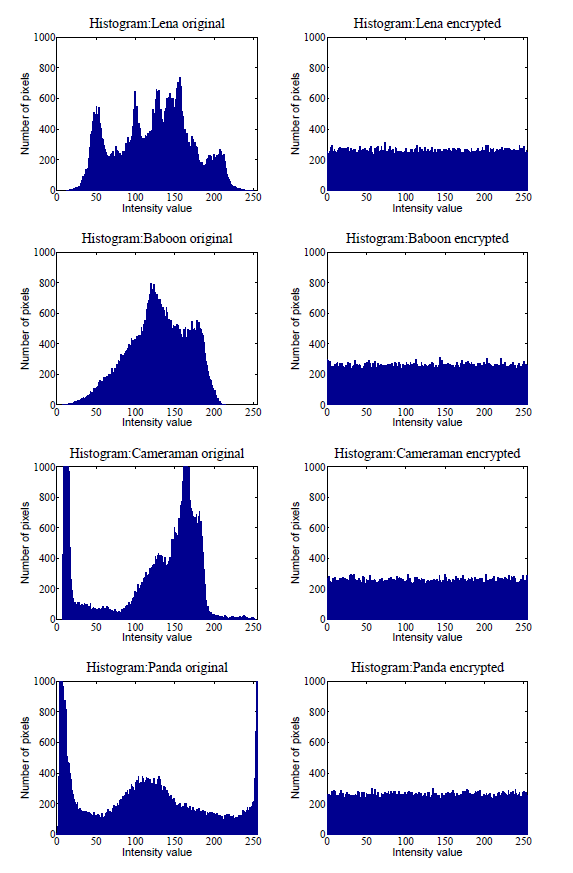}
%
				\caption{Histogram comparison}
		\label{Hist}
	\end{center}
\end{figure}

The Avalanche test of the algorithm shows that a single bit change in key or plain text brings around 49\% change in the cipher bits, which is close to the ideal 50\% change.  The results in Fig. \ref{Images1} show that the accurate decryption is possible only if the correct key is used to decrypt image, else the image remains non recognizable.  For a visual demonstration of avalanche test, the wrong key has a difference of just bit from the original key, the strength of the algorithm can be perceived from this result.  To perform entropy and histogram tests we have chosen five popular 8-bits grey scale images.  Further in the results of histogram in Fig. \ref{Hist} for the original and encrypted image, the uniform distribution of intensities after the encryption is an indication of desired security.  An 8-bits grey scale image can achieve a maximum entropy of 8 bits.  From the results in table \ref{Entropy}, it can be seen that the entropy of all encrypted images is close to maximum, depicting an attribute of the algorithm. 
  
\begin{figure}[h]
	\begin{center}
					\includegraphics[width=8cm]{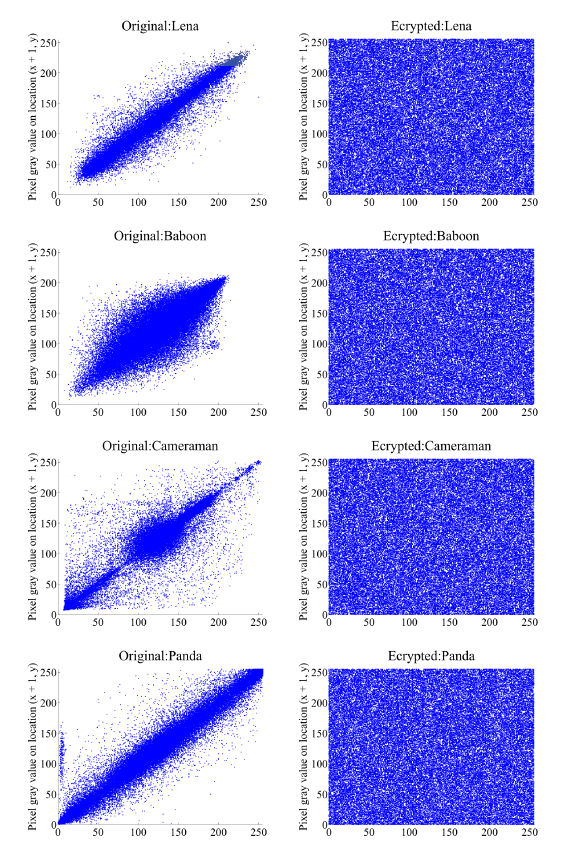} 
				\caption{Correlation comparison}
				\label{CorrComp}
	\end{center}
\end{figure}

Finally the correlation comparison in Fig. \ref{CorrComp} illustrates the contrast between original and encrypted data. Original data, which in our case is an image can be seen to be highly correlated and detaining a high value for correlation coefficient.  Whereas the encrypted image does not seem to have any correlation giving strength to our clause in section \ref{Corel}

\begin{table}[h]
\begin{tabular}{ |p{1cm}|p{1cm}|p{1cm}|p{1cm}|p{1cm}|p{1cm}|  }
 \hline
 \multirow{2}{1cm}{\textbf{Image}}& \multirow{2}{1cm}{\textbf{Size}} &\multicolumn{2}{c|}{\textbf{Correlation}} &\multicolumn{2}{c|}{\textbf{Entropy}}\\\cline{3-6}

&     & Original &Encrypted  &Original &Encrypted  \\
\hline
Lena & 256 x 256 & 0.9744 & 0.0012 &7.4504 & 7.9973\\
\hline
Baboon & 256 x 256 & 0.8198 & 0.0023 &7.2316 & 7.9972\\
\hline
Cameraman& 256 x 256 & 0.9565 & 0.0012 &7.0097 & 7.9973\\
\hline
Panda & 256 x 256 & 0.9811 & 0.0022 &7.4938 & 7.9971\\
\hline
\end{tabular}
\caption{Results for Correlation and Entropy}
\label{Entropy}
\end{table}

\section{Future Work}
For future research, the implementation of the algorithm on hardware and software in various computation and network environment is under consideration. Moreover, the algorithm can be optimized in order to enhance the performance according to different hardware platforms. Hardware like FPGA performs the parallel execution of the code, the implementation of the proposed algorithm on an FPGA is expected to provide high throughput. The scalability of algorithm can be exploited for better security and performance by changing the number of rounds or the architecture to support different key length. 

\section{Conclusion}\label{Conclusion}

In the near future Internet of Things will be an essential element of our daily lives. Numerous energy constrained devices and sensors will continuously be communicating with each other the security of which must not be compromised. For this purpose a lightweight security algorithm is proposed in this paper named as SIT. The implementation show promising results making the algorithm a suitable candidate to be adopted in IoT applications.  In the near future we are interested in the detail performance evaluation and cryptanalysis of this algorithm on different hardware and software platforms for possible attacks.

\bibliographystyle{IEEEtran}
\bibliography{ref1}
\end{document}